\documentclass{article}
\usepackage{epsfig}
\usepackage{amsmath}

\date{\today}

\begin{document}
\title{ Kaluza-Klein black holes with squashed horizons
\\and $d=4$ superposed monopoles}
 \author{{\large {\bf Yves Brihaye}}$^{\dagger}$ and
{\large {\bf Eugen Radu}}$^{\ddagger}$  \\ \\
$^{\dagger}${\small  Physique-Math\'ematique, Universite de
Mons-Hainaut, Mons, Belgium} \\
$^{\ddagger}${\small Department of
Mathematical Physics,  National University of Ireland, Maynooth, Ireland}\\  
 }
\maketitle
 
\begin{abstract}
We present new solutions of the $d=5$ Einstein-Yang-Mills theory
describing black holes with squashed horizons. 
These configurations are asymptotically locally flat and have a
boundary topology of a fibre bundle $R\times S^1 \hookrightarrow S^{2}$.
In a $d=4$ picture, they describe
black hole solutions with both nonabelian and U(1) magnetic charges.
\end{abstract}

\section{Introduction}
The last years have seen an increasing interest in the solutions of 
Einstein equations involving more than four dimensions.
Solutions with a number of compact dimensions, present for
$d\geq 5$ spacetime dimensions, are of particular interest, since they 
exhibit new features that have no analogue in the usual $d=4$ case.
Restricting to the case $d=5$, the simplest configuration of this 
type is found by assuming translational symmetry along the 
extra coordinate direction and corresponds to a uniform vacuum black string
with horizon topology $S^{2}\times S^1$, approaching asymptotically the 
four dimensional Minkowski-space times a circle ${\cal M}^{4}\times S^1$.
The Kaluza-Klein (KK) black hole solutions
\cite{Kudoh:2004hs} have a $S^{3}$ horizon topology
and the same asymptotic structure, presenting a dependence on the compact extra dimension.

However, as shown by the Gross-Perry-Sorkin (GPS)  monopole solution
\cite{Gross:1983hb,Sorkin:1983ns}, there are also $d=5$ 
asymptotically locally flat configurations, approaching a twisted
$S^1$ bundle over a four dimensional Minkowski spacetime.
Black hole solutions with this type of asymptotics enjoyed recently some 
interest, following the discovery by
Ishihara and Matsuno (IM) \cite{Ishihara:2005dp} of a 
new charged solution in the five dimensional Einstein-Maxwell theory. 
The horizon  of the IM black hole has $S^3$ topology, and
its spacelike infinity is a squashed sphere or $S^1$ bundle over $S^2$.
The mass and thermodynamics of this solution  have been discussed in \cite{Cai:2006td}.
A vacuum black hole solution  with similar properties has been presented in \cite{Chen:1999rv}.
As found in \cite{Ishihara:2006iv}, the IM solution admits
multi-black hole generalizations.
KK rotating black hole solutions with squashed horizon
in $d=5$ Einstein gravity are discussed in  \cite{Wang:2006nw}.
Dilatonic generalizations of the IM black holes are studied in \cite{Yazadjiev:2006iv}.

However, these studies restricted to 
the case of an abelian matter content.
At the same time, a number of results obtained in the literature clearly
indicates that the solutions of Einstein's equations coupled to nonabelian
matter fields possess a much richer structure 
than the U(1) counterparts.
In the five dimensional case, we note that
without gravity, the pure Yang-Mills (YM) theory in a flat background
admits topologically stable, particle-like  and
vortex-type solutions obtained by uplifting the $d=4$ YM instantons and
$d=3$ Yang-Mills-Higgs (YMH) monopoles.
However, as found in \cite{Volkov:2001tb},
the particle spectrum become completely destroyed 
by gravity.
Assuming the metric and matter fields to be independent on the extra coordinate,
the Einstein-Yang-Mills (EYM) equations present 
black string solutions \cite{Hartmann:2004tx}.
These configurations approache at infinity
the ${\cal M}^{4}\times S^1$ background and 
possess a nontrivial zero event horizon radius limit
\cite{Volkov:2001tb}.
We mention also the 
 EYM  particle-like and black hole solutions  
 studied in \cite{Brihaye:2002jg}, representing nonabelian generalizations of
the Schwarzschild-Tangherlini solution \cite{tang}.
These configurations are sphericaly symmetric in the  five dimensional spacetime
and are sustained by higher order terms in the YM hierarchy,
approaching at infinity the $d=5$ Minkowski background ${\cal M}^{5}$. 
 
In this paper we study a different type of $d=5$ EYM solutions,
corresponding to black holes with squashed horizons, generalizing for a
SU(2) field the abelian IM solution  \cite{Ishihara:2005dp}.
These solutions have a number of common features with the abelian
counterparts, in particular the event horizons 
is a squashed $S^3$, and the asymptotic structure
is similar to that of the GPS monopole. 
However, they possess a complicated structure, with several distinct
branches and a maximal allowed value of the event horizon radius.
 
In a four dimensional picture, these configurations describe
black hole solutions 
in a Einstein-Yang-Mills-Higgs-dilaton-U(1) (EYMHdU(1)) theory, possesing both
nonabelian and U(1) magnetic charges.

\section{The model}
 \subsection{The action principle }
We consider the EYM-SU(2) action with the Gibbons-Hawking boundary term \cite{Gibbons:1976ue},
\begin{equation}
\label{action5}
I_5=\int_{\mathcal{M}} d^{5}x\sqrt{-g }\big(\frac{R }{16\pi G}
-\frac{1}{2 }{\rm Tr}\{F_{MN }F^{MN}\}\Big)
-\frac{1}{8 \pi G}\int_{\partial\mathcal{M}} d^3 x\sqrt{-h}K,
\end{equation}
where $h_{ij}$ is the induced metric on the boundary $\partial \mathcal{M}$
and $K_{ij}$ is the extrinsic curvature of this boundary, with $K=K_{ij}h^{ij}$.
Apart from gravity, (\ref{action5}) contains an SU(2) gauge field
$A_{M} = \frac{1}{2} \tau_a A_M^{(a)}$, with the field strength tensor 
$F_{MN} =
\partial_M A_N -\partial_N A_M - ig  \left[A_M , A_M \right]$
and gauge coupling constant $g$.

Variation of the action (\ref{action5})
 with respect to  $g^{MN}$  and $A_M$ leads to the field equations
\begin{equation}
\label{field-eqs}
R_{MN}-\frac{1}{2}g_{MN}R   = 8\pi G~T_{MN},~~\nabla_{M}F^{MN}-ig[A_{M},F{MN}]=0.
\end{equation}
where the YM stress-energy tensor is
\begin{eqnarray}
T_{MN} = 2{\rm Tr}
    \{ F_{MP} F_{NO} g^{PO}
   -\frac{1}{4} g_{MN} F_{BC} F^{BC}\}.
\end{eqnarray}
In what follows we will  
assume that both the matter functions and
the metric functions are
independent on the extra coordinate $x^5$.
 Without any loss of generality, we consider a five-dimensional
metric parametrization (with $a=2/\sqrt{3}$)
\begin{eqnarray}
\label{metrica}
ds^2 = e^{-a\psi}\gamma_{\mu \nu}dx^{\mu}dx^{\nu}
 + e^{2a\psi }
 (dx^5 + 2{\cal W}_{\mu}dx^{\mu})^2.
\end{eqnarray}
The four dimensional reduction of the $d=5$ EYM theory with respect to the Killing vector
$\partial/\partial x^5$ has been discussed in \cite{Brihaye:2005pz}, ${\cal W}_\mu$ corresponding
in this picture to a $d=4$ U(1) potential.
For the  reduction of the YM action term,
a convenient SU(2) ansatz is
\begin{eqnarray}
\label{SU2}
A={\cal A}_{\mu}dx^{\mu}+\Phi (dx^5+2 {\cal W}_\mu dx^\mu),
\end{eqnarray}
where  
${\cal A}_{\mu}$ is a purely four-dimensional nonabelian gauge field potential,
while  $\Phi$ corresponds after the dimensional reduction to a
triplet Higgs field.

After Kaluza-Klein reduction with respect to the Killing vector $\partial/\partial x^5$,
we find a $d=4$ gravitating 
YMH model nontrivially interacting with a
dilaton and a U(1) field
\begin{eqnarray}
\label{action4}
& I_4=\int d^{4}x\sqrt{-\gamma }\Big[
\frac{1}{4\pi G}\big(
\frac{\mathcal{R} }{4}
-\frac{1}{2}\nabla_{\mu}\psi \nabla^{\mu}\psi
-\frac{1}{4}e^{3a\psi}G_{\mu \nu }G^{\mu \nu } \big)
-\frac{1}{2}e^{a\psi}\left(Tr\{
{\cal F}'_{\mu \nu }{\cal F}'^{\mu \nu }\}
+e^{-3a\psi}Tr\{ D_{\mu}\Phi D^{\mu}\Phi\}\right)
\Big]_.
\end{eqnarray}
Here $\mathcal{R}$ is the Ricci scalar for the metric $\gamma_{\mu
\nu}$,
while 
 ${\cal F}_{\mu \nu }=
\partial_{\mu}{\cal A}_{\nu}
-\partial_{\nu}{\cal A}_{\mu}-ig[{\cal A}_{\mu},{\cal A}_{\nu}  ]$
,
 $G_{\mu \nu}=\partial_{\mu}{\cal W}_{\nu}-\partial_{\nu}{\cal
W}_{\mu}$
are the SU(2) and U(1) field strength tensors defined in $d=4$
and we note ${\cal F}'_{\mu \nu }={\cal F}_{\mu \nu }+2\Phi G_{\mu \nu}$.
 
 \subsection{The ansatz and field equations}
 
In this paper we consider a $d=5$ metric ansatz used in previous studies on Kaluza-Klein monopoles
\begin{eqnarray}
\label{metrica5d}
ds^2=-L(r)dt^2+U(r)dr^2+B(r)(d \theta^2 + \sin^2 \theta  d \phi^2 )
+F(r)(dx^5+4n\sin^2\frac{\theta}{2}d \varphi)^2,
\end{eqnarray}
Here $\theta$ and $\varphi$ are the standard angles parametrizing 
an $S^2$ with ranges $0 \leq \theta \leq \pi,~ 0 \leq \varphi \leq 2\pi$,
 $n$ being an arbitrary real constant.
 This spacetime has an isometry group $SO(3)\times U(1)$.
 A three dimensional surface $r=const.,t=const.$ has the topology
 of the Hopf bundle, 
$S^1$ fiber over $S^2$ base space.
To avoid a Dirac-Misner string singularity, 
the period of the extra coordinate $x^5$ is restricted 
to $8\pi n$. 

The SU(2) YM ansatz compatible with the symmetries of the above line element reads
\begin{eqnarray}
\label{A5}
A=\frac{1}{2g} \Big\{u(r)\tau_3dt+
  w(r)  \tau_1    d \theta
+\left(\cot \theta \tau_3
+ w(r ) \tau_2  \right) \sin \theta d \varphi
+  H(r)(dx^5 + 4n\sin^2\frac{\theta}{2}d \varphi)\tau_3 \Big\},
\end{eqnarray}
$\tau_a$ corresponding to the Pauli matrices.
In this work we will restrict to a $u(r)=0$ consistent 
truncation of the above ansatz,
the issue of dyonic generalizations being briefly discussed in Section 5.
The equations satisfied by the $d=5$ metric functions and gauge potentials 
are
\begin{eqnarray}
\label{eqs5d}
\nonumber
&  \frac{B'^2}{4B}+\frac{F'}{2F}(B'+\frac{BL'}{4L})+\frac{L'}{2L}
(B'+\frac{BF'}{4F})+U(-1+\frac{n^2F}{B})
-\frac{4\pi G}{g^2}(2w'^2 +\frac{B}{F}H'^2
-\frac{2U}{F}H^2w^2-\frac{U}{B}(w^2-1+2nH)^2)=0,
\\
\nonumber
& F''-\frac{B'}{3B}(\frac{FB'}{2B}-2F'+\frac{FL'}{L})
-\frac{F'}{2}(\frac{U'}{U}+\frac{F'}{F}-\frac{2L'}{3L})
+\frac{2UF(B-7n^2F)}{3B^2}
\\
\nonumber
& -\frac{8\pi G}{g^2}\left(\frac{2F}{3B}w'^2-\frac{5}{3}H'^2-\frac{2U}{B}H^2w^2
+\frac{FU}{B^2}(w^2-1+2nH)^2 \right)=0,
\\
\nonumber
& B''+\frac{B'}{6}(-\frac{3U'}{U}-\frac{B'}{B}+\frac{F'}{F}+\frac{L'}{L})
-\frac{BF'L'}{6FL}+\frac{2U}{3B}(-2B+5n^2F)
+\frac{4\pi G}{g^2}(\frac{8}{3}w'^2-\frac{2B}{3F}H'^2+\frac{2U}{B}(w^2-1+2nH)^2)=0,
\\
\nonumber
& L''-\frac{B'}{3B}(\frac{LB'}{2B}+\frac{LF'}{F}-2L')
-\frac{L'}{2}(\frac{U'}{U}-\frac{2F'}{3F}+\frac{L'}{L})
+\frac{2UL}{3B}-\frac{2n^2UFL}{3B^2}
\\
\nonumber
& -\frac{8\pi G}{g^2}(\frac{2L}{3B}w'^2+\frac{L}{3F}H'^2+\frac{2LU}{BF}H^2w^2
+\frac{UL}{B^2}(w^2-1+2nH)^2)=0,
\\
& w''+w'(\frac{F'}{2F}+\frac{L'}{2L}-\frac{U'}{2U})-w(\frac{U}{F}H^2+\frac{U}{B}(w^2-1+2nH))=0,
\\
\nonumber
& H''-H'(\frac{F'}{2F}-\frac{B'}{B}+\frac{U'}{2U}-\frac{L'}{2L})
-\frac{2nUF}{B^2}(w^2-1+2nH)-\frac{2UH}{B}w^2=0.
\end{eqnarray} 
However, we found more convinient to work with the metric ansatz (\ref{metrica})
 which allows a direct $d=4$ picture, by taking\footnote{The correspondence between the $d=4$ and $d=5$
pictures results straightforward from (\ref{metrica}), (\ref{SU2}). 
One finds $e.g.$  
$L=e^{-a\psi}N\sigma^2,~~F=e^{2a\psi},~~U=e^{-a\psi}/N,~~B=e^{-a\psi}r^2.$}  
\begin{eqnarray}
\label{metrica4}
\gamma_{\mu \nu} dx^{\mu}dx^{\nu}=- N(r) \sigma^2(r) dt^2 + \frac{1}{N(r)} dr^2
+ r^2(d \theta^2 + \sin^2 \theta  d \phi^2 ),
\end{eqnarray}
where
$N(r)=1-2m(r)/r$.
The only nonvanishing component of the $U(1)$ potential is ${\cal W}_{\varphi}=2n\sin^2\theta/2$,
describing a U(1) magnetic monopole.
The $d=4$ YM ansatz is ${\cal A}=\frac{1}{2 } \Big\{ 
  w(r)  \tau_1    d \theta
+\left(\cot \theta \tau_3
+ w(r ) \tau_2  \right) \sin \theta d \varphi\Big\}$, while the Higgs field has only
one component $\Phi=\frac{1}{2}H\tau_3$.
 
For this choice, the equations (\ref{eqs5d}) take a simpler form 
\begin{eqnarray}
\nonumber
&m' =  \frac{1}{2}r^2 N \psi'^2+e^{3a\psi}\frac{n^2}{2r^2} 
 +\frac{4 \pi G}{g^2} \big(e^{ a\psi}(Nw'^2+\frac{(w^2-1+2nH)^2}{2r^2} )
+e^{ -2a\psi}(\frac{1}{2}r^2NH'^2+H^2w^2) \big),
\\
\label{eqs4d}
&\frac{\sigma'}{\sigma}=\frac{8 \pi G}{g^2r}(e^{a\psi}w'^2
+e^{-2a\psi}\frac{r^2H'^2}{2})+r\psi'^2 ,
\\
\nonumber
&(\sigma Nr^2\psi')'=\sigma\bigg\{ \frac{2n^2e^{3a\psi}}{ar^2}+\frac{4 \pi G}{g^2} a\bigg[
e^{ a\psi}(Nw'^2+\frac{(w^2-1+2nH)^2}{2r^2} )
-2e^{ -2a\psi}(\frac{1}{2}r^2NH'^2+H^2w^2)\bigg]\bigg\},
\\
\nonumber
&(e^{a\psi}\sigma Nw')'=\sigma w(\frac{e^{a\psi} (w^2-1+2nH)}{r^2}+e^{-2a\psi}H^2),
\\
\nonumber
&(e^{-2a\psi}\sigma r^2NH')'=2\sigma(e^{-2a\psi}Hw^2+\frac{ ne^{a\psi}(w^2-1+2nH)}{r^2}).
\end{eqnarray}
One can see that for $n\neq 0$, one cannot consistently set $H=0$ unless
$\omega=\pm 1$, $i.e.$ no gauge field.
Thus, it is the Higgs field which 
supports the interaction of the four dimensional YM field with 
the U(1) monopole.

\subsection{Known solutions}
Solutions of the equations (\ref{eqs5d})
are already known in a few particular cases. 
The vacuum
black version of the GPS monopole presented in \cite{Chen:1999rv} 
is found for a pure gauge nonabelian configuration $w(r)=\pm 1,~~~ H(r)=0$
and has a metric form
\begin{eqnarray}
\label{vac1}
ds^2=-(1-\frac{2S}{r})dt^2+(1+\frac{2p}{r})(\frac{dr^2}{1-\frac{2S}{r}}+
r^2(d \theta^2 + \sin^2 \theta d \varphi^2))
+\frac{1}{1+\frac{2p}{r}}(dx^5+4n\sin^2\frac{\theta}{2}d \varphi)^2,
\end{eqnarray} 
where $p=-S/2+\sqrt{n^2+S^2/4}$, the GPS monopole solution
being recovered for $S=0$.
After KK reduction with respect to the Killing vector $\partial/\partial x^5$,
one finds $d=4$ magnetically charged black hole solutions.

The embedded U(1) IM configuration  \cite{Ishihara:2005dp} corresponds  to 
\begin{eqnarray}
\nonumber
 ds^2=\frac{r(r+r_0)}{(r-r_-)(r-r_+)}dr^2+
r(r+r_0)(d \theta^2 + \sin^2 \theta d \varphi^2)-\frac{(r-r_-)(r-r_+)}{r^2} dt^2
\\
+\frac{r(r+r_-)(r+r_+)}{4n^2(r+r_0)}(dx^5+4n\sin^2\frac{\theta}{2}d \varphi)^2,
~~~ A=\pm \frac{1}{4r}\sqrt{\frac{ 3r_+r_-}{\pi G}} \tau_3dt.
\end{eqnarray} 
It may be worth noting that when viewed in a four dimensional
perspective, the IM solution describes black holes
in a theory with two 
U(1) fields which couple different to the 
dilaton\footnote{The corresponding $d=4$ action principle can be 
read from (\ref{action4}) by taking $\Phi=0$ and an abelian
subgroup of SU(2).}.
One of these fields  has an electric charge and the other corresponds to a Dirac monopole.

For $n=0$, the system (\ref{eqs5d}) presents 
nonabelian black string solutions 
originally found in \cite{Hartmann:2004tx}.
In the limit of zero event horizon radius, the 
 nonabelian vortices discussed in \cite{Volkov:2001tb} are recovered.
No exact solutions with reasonable asymptotics
are available analytically in this case and the field equations
have to be solved numerically.
A detailed analysis of the properties of the black string solutions have been presented in 
\cite{Brihaye:2005tx}.
It has been found that the pattern of solutions is very similar to that
observed for non-abelian vortices,
depending crucially on the value of the
gravitational coupling constant  $\alpha=\sqrt{4\pi G }H_0/g$ (with
$H_0$ the asymptotic value of the $H-$function). 
Several branches of solutions exist
and the extend of the branches in $\alpha$ gets smaller and smaller for successive branches.

\subsection{Boundary conditions}
In this work we will consider black hole solutions of the system (\ref{eqs5d}), with
an event horizon located at $r=r_h>0$.
As $r\to r_h$, we have $N(r_h)=0$, while the other functions stay finite and nonzero.
The formal power series expansion near the event horizon in terms of 
the functions which enters the $d=4$ picture is
\begin{eqnarray}
\nonumber
m(r)&=&\frac{r_h}{2}+m'_h(r-r_h)+O(r-r_h)^2,
~~
\sigma(r)=\sigma_h+\sigma'_h(r-r_h)+O(r-r_h)^2,
\\
\label{a1}
\psi(r)&=&\psi_h+\psi'_h(r-r_h)+O(r-r_h)^2,
~~
H(r)=H_h+H'_h(r-r_h)+O(r-r_h)^2,
\\
\nonumber
w(r)&=&w_h+w'_h(r-r_h)+O(r-r_h)^2,
\end{eqnarray}
where 
\begin{eqnarray}
\nonumber
m'_h&=&\frac{n^2 e^{3a\psi_h}}{2r_h^2}+
\frac{4 \pi G}{g^2}\left(e^{a\psi_h}\frac{(w_h^2-1+2nH_h)^2}{2r_h^2}+e^{-2a\psi_h}H_h^2w_h^2\right),
\\
\psi_h'&=&\frac{1}{r_h^2 N_h'}\bigg\{
\frac{2n^2 e^{3a\psi_h}}{ar_h^2}+a \frac{4 \pi G}{g^2}\bigg(e^{a\psi_h}\frac{n(w_h^2-1+2nH_h)^2}{2r_h^2}
-2e^{-2a\psi_h}H_h^2w_h^2\bigg)\bigg \},
\\
\nonumber
w_h'&=&\frac{w_h}{N_h'}\bigg(\frac{(w_h^2-1+2nH_h)}{r_h^2}+e^{-3a\psi_h}H_h^2\bigg),
~~~
H_h'=\frac{2}{r_h^2N_h'}\bigg(H_hw_h^2+e^{3a\psi_h}\frac{n(w_h^2-1+2nH_h)}{r_h^2} \bigg),
\\
\nonumber
\sigma_h'&=&\sigma_h\left( r_h\psi_h'^2+\frac{8 \pi G}{g^2r_h}(e^{a\psi_h}w_h'^2
+e^{-2a\psi_h}\frac{r_h^2H_h'^2}{2})\right),
\end{eqnarray}
with $\sigma_h,~w_h,~H_h, ~~\psi_h$ being arbitrary parameters,
while $N'_h=(1-2m'_h)/r_h$.

We are interested in solutions of the $d=5$ EYM equations whose metric asymptotic structure
is similar to that of the GPS monopole,  $i.e.$  with a line element 
 as $r \to \infty$  
\begin{eqnarray}
\label{asy1}
ds^2=-dt^2+dr^2+r^2(d \theta^2 + \sin^2 \theta  d \phi^2 )
+ (dx^5+4n\sin^2\frac{\theta}{2}d \varphi)^2.
\end{eqnarray} 
The analysis of the field equations as $r\to\infty$ for this metric
asymptotics gives 
\begin{eqnarray}
\label{a2}
m(r)=M+\frac{m_1}{r}+\dots,~~\sigma(r)=1+\frac{s_2}{r^2}+\dots,
~~~\psi(r)= \frac{\psi_1}{r}+\frac{\psi_2}{r^2}+\dots,
\\
\nonumber
H(r)=H_0+\frac{h_1}{r}+\frac{h_2}{r^2}+\dots, ~~
w(r)=c_2e^{-H_0 r}+\dots,
\end{eqnarray}
where
\begin{eqnarray}
\nonumber
\psi_2=\frac{3}{4}an^2+\psi_1M+ \frac{\pi aG}{g^2}((1-2H_0n)^2-2h_1^2),
~~~
h_2= h_1(a\psi_1+M)+n(2H_0n-1),
\\
\nonumber
m_1=-\frac{1}{2}(\psi_1^2+ n^2)-\frac{2a\pi G}{g^2}(h_1^2+(1-2H_0n)^2),
~~~s_2=-\frac{\phi_1^2}{2}- \frac{2\pi G}{g^2}h_1^2,
\end{eqnarray}
with $M,~\psi_1,~H_0,~h_1$ and $c_2$  real constants.
 
After dimensional reduction, the $d=5$ EYM 
configurations become four dimensional spherically symmetric black hole 
solutions of the EYMHdU(1) model (\ref{action4}), the event horizon still being
located at $r=r_h$. 
 They will possess a unit nonabelian magnetic charge
and a U(1) magnetic charge  $Q_m=2n$, approaching asymptotically the ${\cal M}^{4}$ background.
The constant $M$ in (\ref{a2}) corresponds to the total
mass of the $d=4$ configurations, while $\psi_1$ gives the dilaton charge.

\section{Numerical solutions}
 Although an analytic or approximate solution of the equations (\ref{eqs4d}) 
 appears to be
intractable, here we present   arguments 
for the existence of nontrivial solutions
which smoothly interpolate between the asymptotic expansions (\ref{a1}), (\ref{a2}).
We restrict our integration to the region outside the event 
horizon\footnote{To integrate the equations, we used the differential
equation solver COLSYS which involves a Newton-Raphson method
\cite{COLSYS}.}.
  $n\neq 0$ generalizations are found for any black string EYM  
configuration by slowly increasing the
parameter $n$ 
(since the transformation $n\to -n$ leaves the field equations unchanged  
except for the sign of the $H$,
we consider here only positive values of $n$).

To find 
dimensionless quantities with the right asymptotics, 
we use the observation that the equations
are left invariant by
the following rescaling
\begin{eqnarray}
\label{resc} 
r \to e^{a\phi_0/2} H_0~r, ~~H \to e^{a\phi_0}H/H_0,
~~m \to m~e^{a\phi_0/2} H_0,~~n \to n~e^{-a\phi_0} H_0.
\end{eqnarray}
where $H_0$ is the asymptotic value of $H(r)$
and $\phi_0$ an arbitrary constant.
Thus, similar to the black string case,  
the constant of the theory are absorbed into the 
coupling constant $\alpha=\sqrt{4 \pi G}H_0/g$, which is an input parameter.
One can take advantage of the this double
rescaling and set  in the numerical analysis
 $H(\infty)=1$ and $\psi(\infty)=0$ without loosing
any generality.

As expected, these configurations have many features in
common with the black string solutions
discussed in \cite{Hartmann:2004tx}; they also present new features that
we will  pointed out in the discussion.

All solutions we found have a positive asymptotic value of the metric function
$m(r)$.
The gauge functions $\omega(r)$ and $H(r)$ interpolates monotonically 
between some constant values on the event 
horizon and zero respectively one at infinity,
without presenting any local extremum (see Figure 1).
For all $r_h>0$, the dilaton function takes a finite  
value at the event horizon.
The profiles of two typical profiles are presented in Figure 1.

\newpage
\setlength{\unitlength}{1cm}

\begin{picture}(18,7)
\centering
\put(2,0.0){\epsfig{file=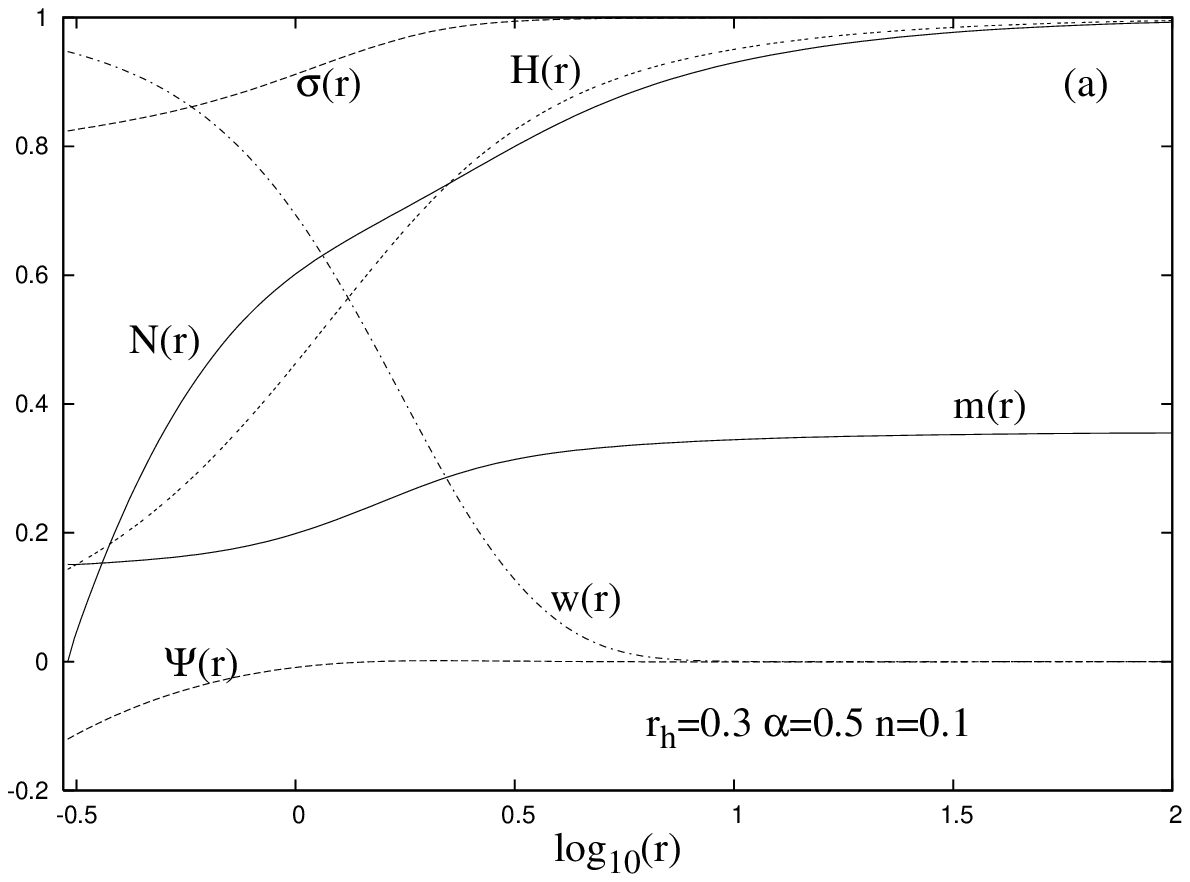,width=12cm}}
\end{picture}
\begin{picture}(19,9.)
\centering
\put(2.6,0.0){\epsfig{file=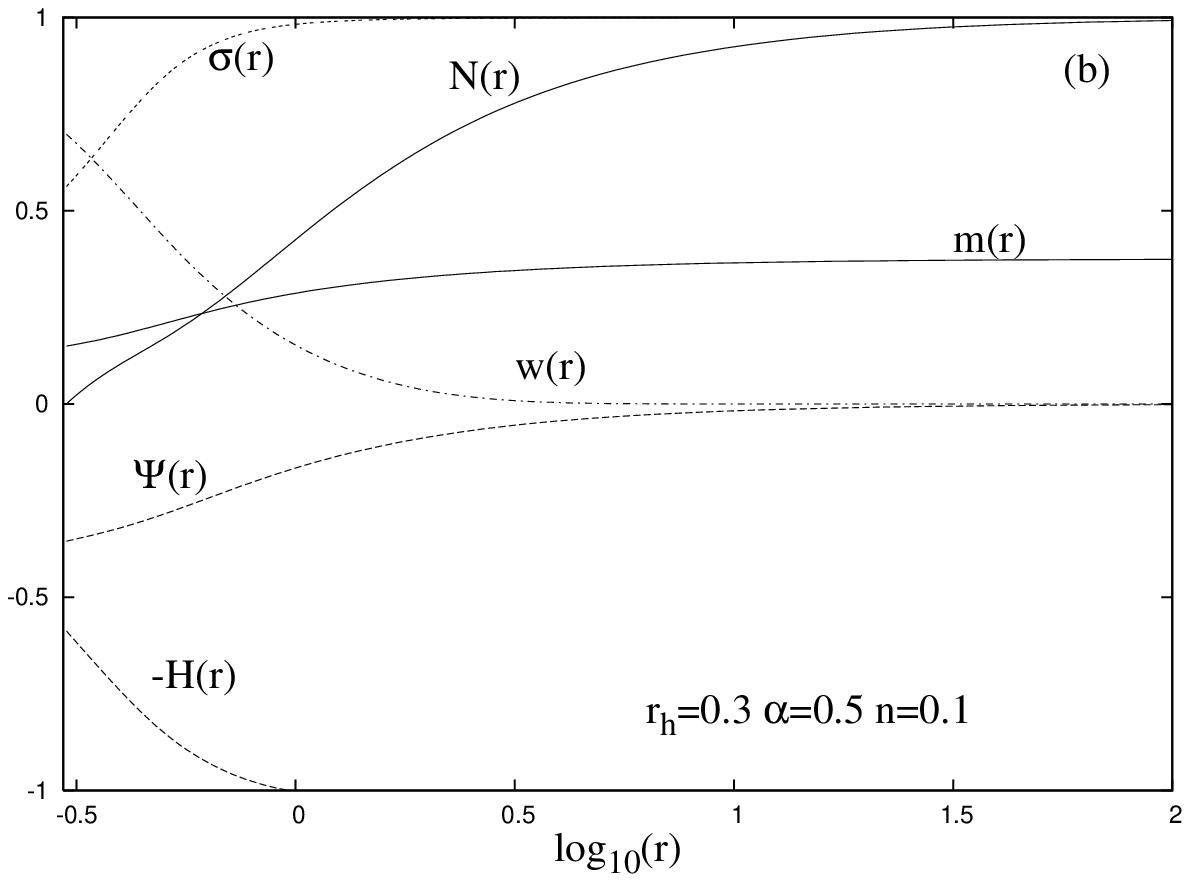,width=12cm}}
\end{picture}
\\
\\
{\small {\bf Figure 1.}
The profiles of the functions $N,m,\sigma,w,H,\psi$ are represented
for typical first (Figure 1a) and second branch (Figure 1b) solutions with
$\alpha=0.5,~r_h=0.3,~n=0.1$.
 }
\\
\\
In the near-horizon region, the area of the squashed spheres is 
proportional to $r^3$.
Thus, similar to the IM case, there is a region
such that the solution will behave as a five-dimensional black hole
for observers in that region.
Far away from the event horizon, the metric function
$g_{55}\equiv e^{2a\psi}$
is almost constant and the spacetime is effectively four dimensional.

The complete classification of the solutions in the
space of parameters $(\alpha,~n,~r_h)$
is a considerable task which is not aimed in this paper.
Instead, we analyzed
in details a few particular classes of solutions which, hopefully,
reflect all relevant properties of the general pattern.
It is important to remember
that for $n=0$ the EYM equations admit several branches of
black string solutions, the number of them depending on $r_h$
and of $\alpha$ \cite{Hartmann:2004tx}.

For simplicity, we have studied mainly solutions
in the region
$\alpha=0.5$ where it is known
that two branches 

\newpage
\setlength{\unitlength}{1cm}
\begin{picture}(18,7)
\centering
\put(2,0.0){\epsfig{file=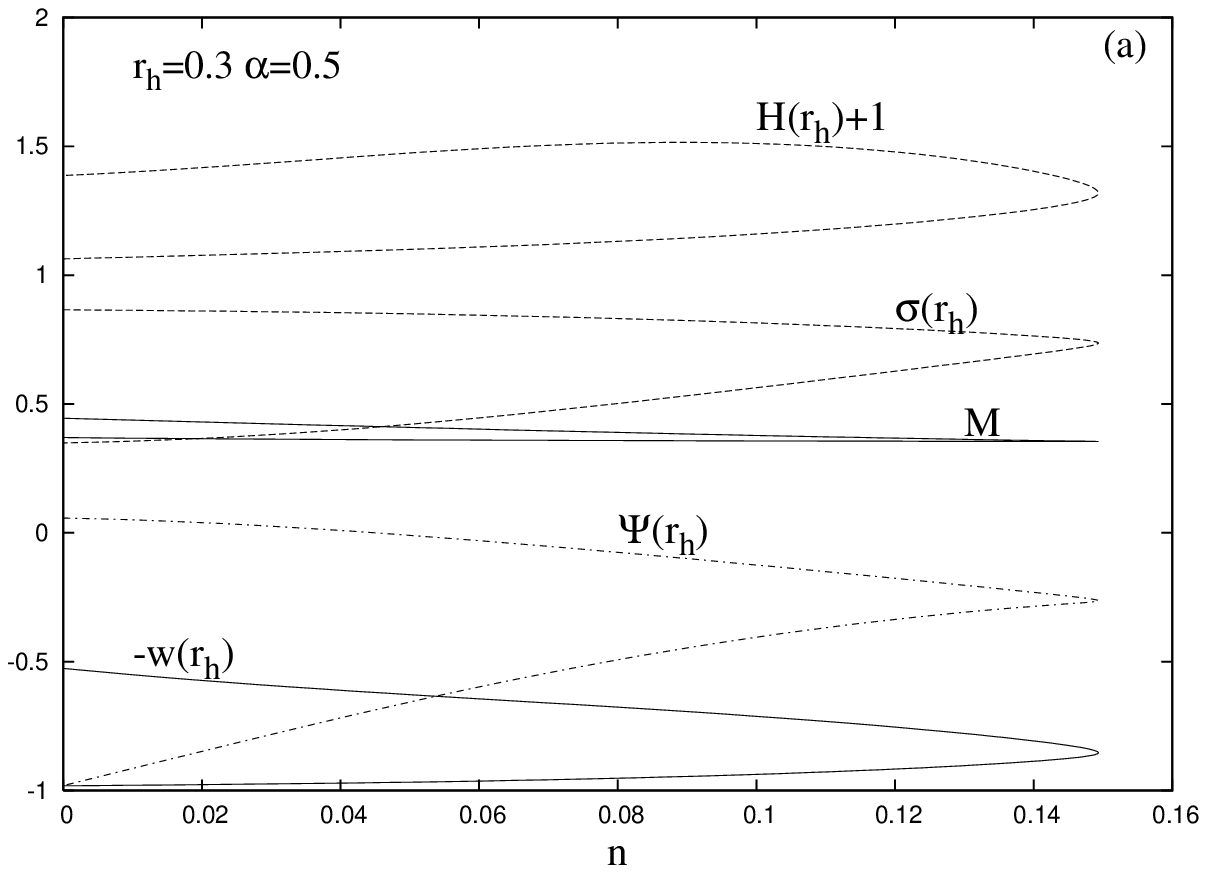,width=12cm}}
\end{picture}
\begin{picture}(19,9.)
\centering
\put(2.6,0.0){\epsfig{file=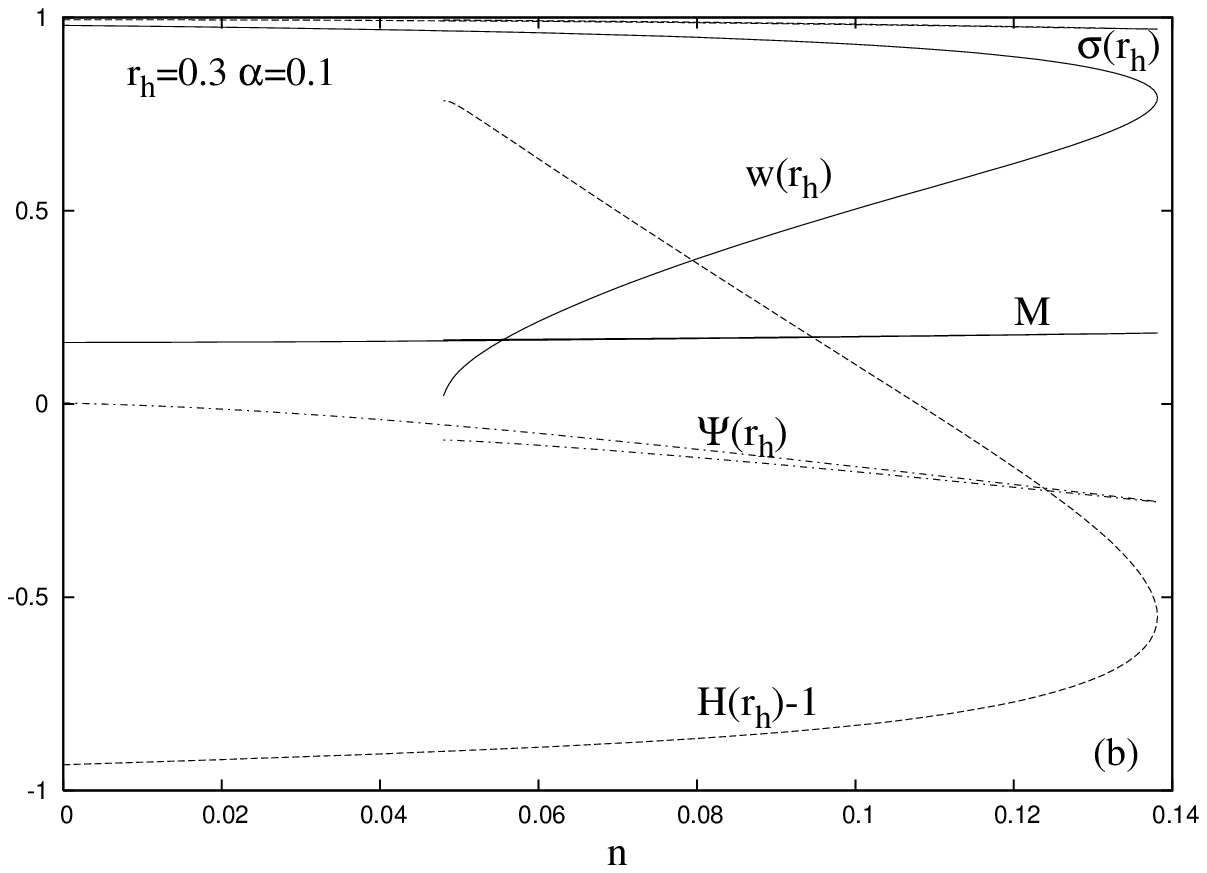,width=12cm}}
\end{picture}

{\small {\bf Figure 2.}
Several relevant quantities are plotted as a function of $n$
for $r_h=0.3$ and two different values of $\alpha$.}
\\
\\
 of black string solutions exist.
 The  $\alpha=0.5$ branches of $n=0$ black string solutions  exist
for $0<r_h< r_{h,max}$ with $r_{h,max} \approx 0.65$. In the limit
$r_h \to 0$, the solutions converge to regular solutions
 first constructed in \cite{Volkov:2001tb}. The regular solutions
 have $N(0)=1$ so that the convergence is not pointlike at $r=0$.
 Accordingly, the surface gravity $\kappa$ of the black 
string (given by
 $\kappa  = \sigma(r_h)N'(r_h)/2$) becomes infinite
 for  $r_h \to 0$; all other parameters,
$e.g.$
 the dilaton and its derivative remain finite.
We expect the features of $n \neq 0$ solutions we will
discuss for $\alpha = 0.5$ to be qualitatively the same when more
than two branches of solutions occur.

We have tried to understand the domain of existence of these
solutions when the  parameter $n$ varies while the other parameters
are kept fixed. It turns out that (at least in the region
for $\alpha=0.5$) the two  black string solutions
existing for $n=0$ get slowly deformed for $n>0$, forming two
different branches extending in $n$.
These two distinct
branches  join at a maximal value of $n$.
Our numerical analysis, strongly suggests that there are no solutions
for $n$ larger than this maximal value.
This phenomenon is illustrated on Figure 2a where a few parameters
characterizing these solutions are reported as functions of  $n$.
With this choice of  

\newpage
\setlength{\unitlength}{1cm}
\begin{picture}(18,7)
\centering
\put(2,0.0){\epsfig{file=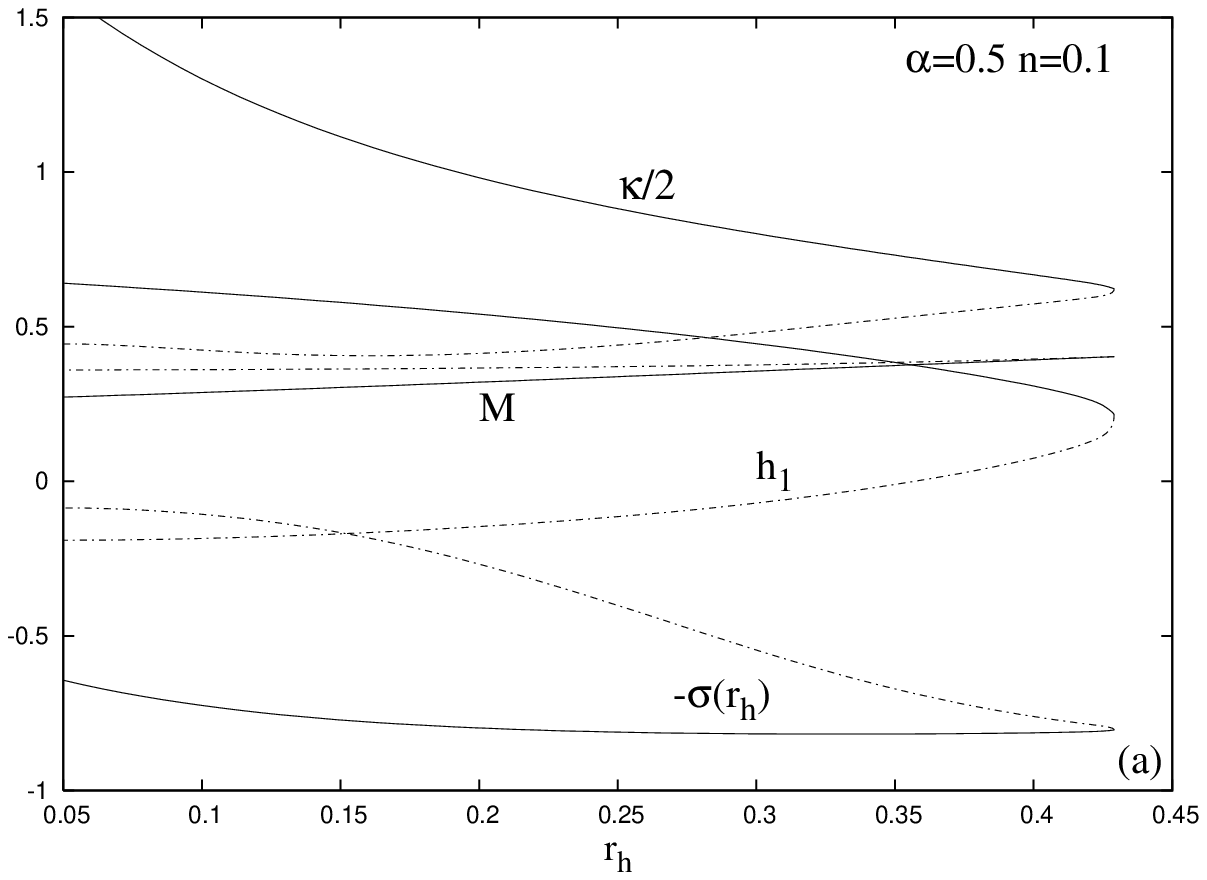,width=12cm}}
\end{picture}
\begin{picture}(19,9.)
\centering
\put(2.6,0.0){\epsfig{file=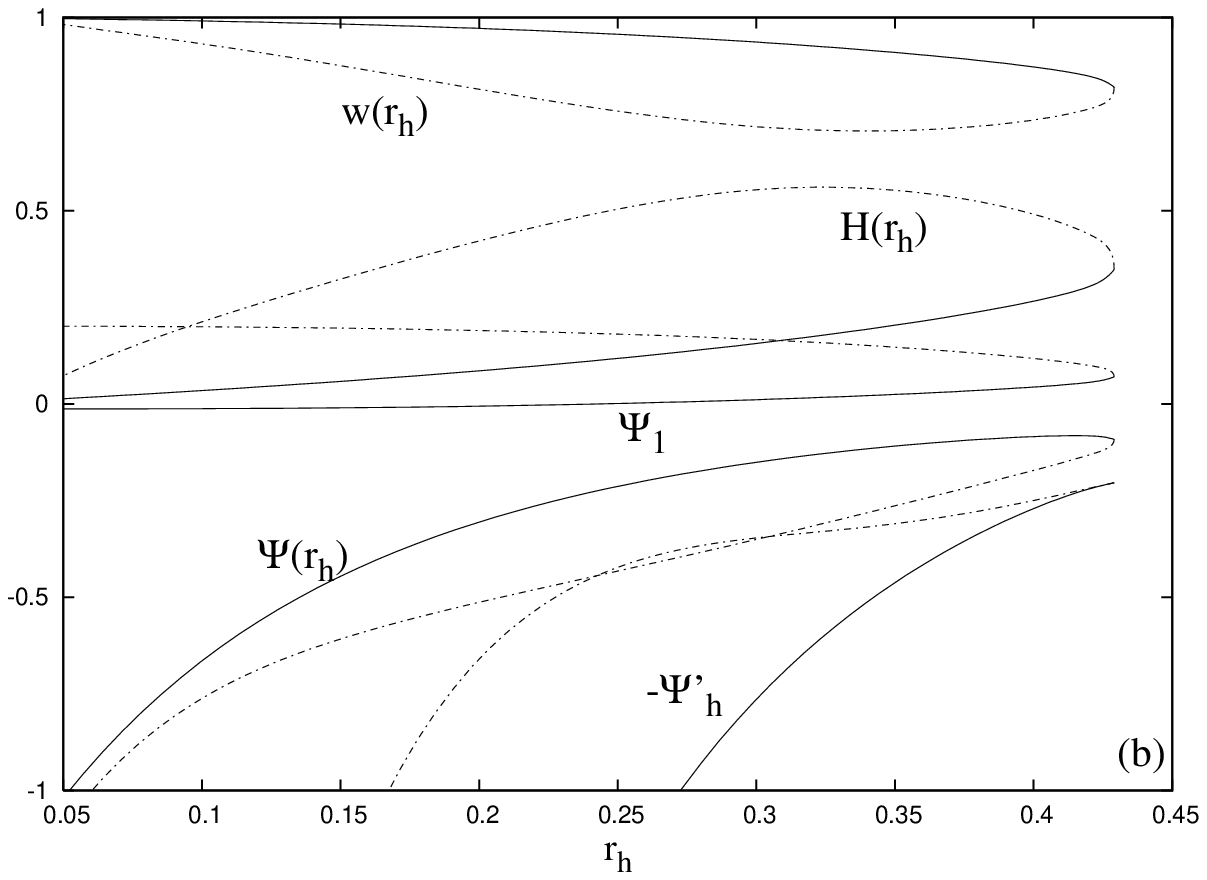,width=12cm}}
\end{picture}
\\
\\
{\small {\bf Figure 3.}
 The dependence
of solution properties on the value of the
 event horizon radius $r_h$ is plotted for $\alpha=0.5,~n=0.1$.
The dotted line indicates the higher mass branch solutions.
$\kappa  = \sigma(r_h)N'(r_h)/2$ is the surface gravity of the
black hole solutions.}
\\
\\
the parameters
 ($\alpha=0.5,~r_h=0.3$)
the two branches terminate at $n \simeq 0.1515$.
 
However, for small values
of $\alpha$ (typically $\alpha < 0.25$) only one black string exists
for a given $r_h$ \cite{Hartmann:2004tx}.
The numerical analysis shows that these black strings are
deformed by $n$, forming a branch which terminates at a maximal value
of $n$ ($e.g.$ $n\simeq 0.138$ for $\alpha=0.1$ and $r_h = 0.3$).
Then we manage to construct a second branch of solutions, terminating
at the same value, say $n=n_{max}$. This second branch, however 
exists only on an interval $n \in [n_{cr},n_{max}]$ 
(with $n_{cr}\approx 0.05$ in our case, see Figure 2b).
In the limit
$n \to n_{cr}$, the solution converges to a configuration 
with $w(r)=0$, $i.e.$ the Yang-Mills field become a Wu-Yang monopole.
The function $H(r)$ remains non trivial and, 
surprisingly, it has $H(r_h) > 1$ and 
decreases  monotonically to its asymptotic value 
$H(\infty)=1$. The second branch solutions have no counterparts in the $n=0$ case.
 
Different from the U(1) case, the EYM 
configurations exist for a limited range
of the event horizon radius only.
The analysis of the behaviour of the solutions in the limit  $r_h \to 0$
is clearly interesting. For $n>0$, the situation
  deeply contrasts from the case $n=0$, where a $d=4$ regular limiting solution
  is available. Indeed, our numerical analysis strongly suggest that
   in the limit $r_h \to 0$ the  $d=4$ black hole solution converge
   to a singular configuration characterized by a diverging value of $\psi(0)$
  and of $\psi'(0)$.
  This agrees with the physical intuition that no nonsingular
particle-like Dirac monopole
can exist.
However, these configurations are globally
regular in a five-dimensional picture, describing SU(2) generalizations
of the  GPS monopoles.
$r=0$ corresponds here to the origin of the coordinate system
and is a  regular point.

These features are illustrated on Figure 3
where the parameters $\alpha$ and $n$ are fixed,
 while the horizon radius $r_h$ is varied. 
The occurence of two branches
of solutions terminating at a maximal value of $r_h$ is clearly visible
on that plot; in this case we have $r_{h,max} \approx 0.429$.
The numerical construction of the solutions
turns out to be difficult for small values of $r_h$ but the results
reported on Figure 3 suggest that
$w(r_h)\to 1$ and $H(r_h)\to 0$ in the limit $r_h\to 0$ for both branches.
This indicates that the non Abelian character of the solution
persists in the  $r_h\to 0$ limit.  
 
The limiting $d=5$ particle-like solutions are found by solving
directly the system (\ref{eqs5d}),
with the boundary conditions $L(0)=L_0>0,~U(0)=1,~F(0)\sim O(r^2)$,
where we fix the metric gauge
by taking $B(r)=r^2$.
The boundary conditions satisfied by gauge field
potentials are $w(0)=1$, $H(0)=0$.
The asymptotics at infinity of the particle-like
solutions are similar to those of the
black hole counterparts. This behaviour strongly contrasts with that of the
globally regular vortices of \cite{Volkov:2001tb}
or the particle-like solutions in
\cite{Brihaye:2002jg}.
A systematic study of the  particle-like solutions emerging
in the $r_h\to 0$ limit
 will be presented elsewhere.
\section{The mass of the $d=5$ solutions}
The construction of the conserved quantities for solutions
with the asymptotic structure (\ref{asy1})
is an interesting problem.
At a conceptual level, the background subtraction method is not
entirely satisfactorly, since it relies on
the introduction of a spacetime which is auxiliarly to the problem
and is not obvious in this case.

Inspired by the AdS results \cite{Balasubramanian:1999re},
Kraus {\it et al.} have proposed in \cite{Kraus:1999di} a counterterm in
a five-dimensional asymptotically flat spacetime with boundary
topology $R\times S^{3}$ or $R^{2}\times S^{2} $.
By taking the
variation of the action  plus this counterterm
part with respect to the boundary metric $h_{ij}$, one  finds the
boundary stress-energy tensor \cite{Brown:1992br}, and then define the
conserved charge  associated with some Killing vector of the boundary metric
(see \cite{Astefanesei:2005ad}
for further work in this direction). 

The computation of the action and conserved charges of a Kaluza-Klein
monopole within this approach has been done by Mann and Stelea \cite{Mann:2005cx},
who have proposed a  counterterm 
expression\footnote{In this Section we do not take the rescaling (\ref{resc}).} 
\begin{equation}
I_{ct}=\frac{1}{8\pi G}\int d^{4}x\sqrt{-h}\sqrt{2\mathsf{R}} \,
, \label{countertermaction}
\end{equation}
where $\mathsf{R}$ is the Ricci scalar of the induced metric on
the boundary.  With this counterterm, the  boundary
stress-energy tensor is found to be
\begin{equation}
\label{mann} T_{ij}=\frac{1}{8\pi G}\left( K_{ij}-Kh_{ij}-\Psi(
\mathsf{R}_{ij}-\mathsf{R}h_{ij})-h_{ij}h^{kl}\Psi_{;kl}+\Psi_{;ij}
\right)\,,
\end{equation}
where $K$ is the trace of extrinsic curvature $K_{ij}$ of the
boundary, and $\Psi=\sqrt{2/{\mathsf{R}}}$. If the boundary
geometry has an isometry generated by a Killing vector $\xi ^{i}$,
then $ T_{ij}\xi ^{j}$ is divergence free, 
from which it follows
that the quantity
\begin{equation}
\mathcal{Q}=\int_{\Sigma }d\Sigma_{i}T^{i}{}_{j}\xi ^{j},
\label{concharge}
\end{equation}
associated with a closed surface $\Sigma $, is conserved. 

Using the asymptotic expression (\ref{a2}) for the metric functions 
$m(r),~\sigma(r)$ and $\psi(r)$ we 
find the boundary
stress-energy tensor
of the EYM solutions\footnote{Note that a different counterterm choice leads
to a different expression of the $T_{\theta }^{\theta },~T_{\varphi }^{\varphi } $
components of the boundary stress tensor, but the same action and conserved charges.}
\begin{eqnarray}
8\pi GT_{~t}^{t} &=&\frac{2M}{r^{2}}+O(1/r^{3}), ~~~
8\pi GT_{~5 }^{5 } =\frac{2M-3a\psi_1}{2r^{2}}+O(1/r^{3}),
\label{btik}
\\
8\pi GT_{\varphi }^{\varphi }&=&
8\pi GT_{\theta }^{\theta } =-\frac{ M+2m_1+n^2+4s_2}{2r^{3}}+O(1/r^{4}),
~~~8\pi GT_{~\varphi }^{5 } =\frac{2n(2M-3a\psi_1) \sin^2\theta/2 }{r^{2}}+O(1/r^{3}).
\notag
\end{eqnarray} 
The solutions' mass is 
the conserved charge associated with the Killing vector
$\partial/\partial t$ of the boundary metric
\begin{equation}
\mathcal{M}=\frac{8\pi Mn}{G}.
\end{equation}  
As usual, a positive-definite metric is found by
considering in (\ref{metrica5d}) the analytical continuation 
$t\rightarrow i t$.
In this case, the absence of conical singularities at the root $%
r_{h}$ of the function $N(r)$ imposes a periodicity 
in the Euclidean time coordinate 
\begin{equation}
 \label{new-rel}
\beta =\frac{4\pi }{N^{\prime }(r_{h})\sigma(r_h)}~, 
\end{equation}%
the  Hawking temperature being $T_H=1/\beta$.

One can also prove that although the horizon of these black holes is deformed,
their entropy still obeys the area formula.
One starts by evaluating the classical tree-level action $I_5$ \cite{Gibbons:1976ue},
where the $R$ volume term is replaced with
$2R_t^t-16\pi G T_t^t$.
For our purely magnetic ansatz, the term $T_t^t$  exactly cancels the matter field
lagrangean in the bulk action
 $L_m=-1/2g^2Tr(F_{MN }F^{MN})$ (see also the general discussion in \cite{Visser:1993qa}).
To evaluate the integral of $R_t^t$ one uses the Killing 
identity $\nabla^M\nabla_N  \zeta_M=R_{N M}\zeta^M,$
for the Killing vector $\zeta^M=\delta^M_t$.
As expected, the counterterm action (\ref{countertermaction})
regularizes also the action (\ref{action5}) and, upon application 
of the Gibbs-Duhem relation $S=\beta \mathcal{M}-I_5$
we find the entropy 
$
S=8\pi^2 n r_h^2,
$
which is one quarter of the event horizon area.
\section{Further remarks}
Black objects in $d=5$ dimensions have a much richer spectrum 
of horizon topology than the four dimensional solutions.
The black hole solutions with an $S^3$ horizon topology
and approaching asymptotically a twisted
$S^1$ bundle over a four dimensional Minkowski spacetime
are a particularly interesting case.
For such solutions, the spacetime behaves as a five-dimensional black hole
near the horizon, while the dimensional reduction to four is realized 
in the far region.

In this work we have analysed the basic properties of this type of 
solutions in EYM-SU(2) theory.
We have found that despite the existence of 
a number of similarities to the U(1) IM solution,
the nonabelian configurations exhibit some new qualitative features,
in particular a complicated branch structure
and a different zero event horizon radius limit.
>From a four dimensional perpective, these solutions correspond to
dilatonic-Reissner-Nordstr\"om black holes
sitting inside the center of a nonabelian monopole.

It would be interesting to consider the axially symmetric
generalizations of these configurations, with a $d=5$ ansatz
presenting a nontrivial dependence on the azimuthal coordinate $\theta$.
The static $d=4$ solutions will also be axially symmetric, with
a U(1) field possessing both a U(1) magnetic monopole charge and a magnetic dipole moment.

Following \cite{Brihaye:2005tx},
new interesting $d=4$ solutions can be found 
by boosting the $d=5$ solutions in the $(x^5,t)$-plane,
$x^5=\cosh \beta~U+\sinh \beta~T,~~t=\sinh \beta~U+\cosh \beta~T$, 
where $\beta$ is an arbitrary parameter.
The dimensional reduction of the $d=5$ EYM configurations along the
$U$-direction provides new solutions of the EYMHd-U(1) model (\ref{action4}) \cite{Brihaye:2005tx}.
The resulting $d=4$ line element reads 
\begin{eqnarray}
\label{new4D}
 \bar{\gamma}_{\mu \nu}dx^{\mu}dx^{\nu}=
  -e^{a(\psi-\bar{\psi})}N\sigma^2
  (dT-4n\sinh \beta\sin^2\frac{\theta}{2}
d \varphi)^2+e^{a(\bar{\psi}-\psi)}\left(\frac{dr^2}{N}+r^2
(d \theta^2 + \sin^2 \theta  d \phi^2)\right),
\end{eqnarray}
where $\bar{\psi}=\psi+\frac{1}{2a}\log (\cosh^2 \beta
-\sinh^2 \beta e^{-3a\psi} N\sigma^2)$ is the new dilaton field.
The new $U(1)$ field has both electric and magnetic components
\begin{eqnarray}
\label{new4D-W}
 \bar{{\cal W}} =\frac{2n e^{2a\psi}\cosh \beta~\sin^2\frac{\theta}{2}}
{e^{2a\psi}\cosh^2 \beta-e^{-a\psi}N\sigma^2\sinh^2 \beta}d\varphi+
 \frac{1}{2}
\frac{(e^{2a\psi}-e^{-a\psi}N\sigma^2)\sinh \beta \cosh \beta}
{e^{2a\psi}\cosh^2 \beta-e^{-a\psi}N\sigma^2\sinh^2 \beta}d T.
\end{eqnarray}
The $ {\cal A}_r$ and ${\cal A}_{\theta}$ components of the $d=4$ SU(2) gauge field 
are not affected, while 
\begin{eqnarray}
\label{new4D-YM}
\bar{A_\varphi}=A_\varphi-4\Phi n \sin^2\frac{\theta}{2}
\frac{e^{-2a\psi}N\sigma^2\sinh^2 \beta}{e^{2a \psi}\cosh^2 \beta-e^{-a\psi}N\sigma^2 \sinh^2 \beta}~,
~~
 \bar{A_T}=\Phi \sinh \beta 
\frac{e^{-a\psi}N\sigma^2}{e^{2a \psi}\cosh^2 \beta-e^{-a\psi}N\sigma^2 \sinh^2 \beta},
\end{eqnarray}
the new Higgs field being $\bar{\Phi}=\Phi \cosh \beta$.
Different from the seed solution,
these configurations possess a nut-charge $\bar{n}=n\sinh \beta $,
and are dilatonic-U(1) generalizations of the "nutty dyons" discussed in 
\cite{Brihaye:2005ak}, representing the nonabelian version of
the Brill solution \cite{Brill}.
The gauge fields possess in this case both electric and magnetic charges.

A systematic discussion of these aspects, 
together with higher winding number generalizations of the  black hole solutions
and particle-like configurations
will be presented elsewhere.
\\
\\
{\bf\large Acknowledgements} \\
The authors thank  Cristian Stelea for  valuable remarks on a
draft of this paper.
YB is grateful to the
Belgian FNRS for financial support.
The work of ER is carried out
in the framework of Enterprise--Ireland Basic Science Research Project
SC/2003/390 of Enterprise-Ireland.


\end{document}